\documentclass[twocolumn,aps,showpacs,floatfix,prc]{revtex4}
\usepackage[dvips]{epsfig}
\begin{document}

\title{ Energy Dependence of exotic nuclei production cross sections by photofission reaction in GDR range }
\author{ Debasis Bhowmick$^{1*}$, F. A. Khan$^{2\S*}$, Debasis Atta$^{3\dagger*}$, D. N. Basu$^{4*}$ and Alok Chakrabarti$^{5*}$}

\address{ $^*$ Variable  Energy  Cyclotron  Centre, 1/AF Bidhan Nagar, Kolkata 700 064, India}
\address{ $^{\S}$ Dept. of Physics, University of Kashmir, Hazratbal, Srinagar 190 006, India}
\address{ $^\dagger$ Shahid Matangini Hazra Govt. Degree College for Women, Tamluk, West Bengal 721 649, India}

\email[E-mail 1: ]{dbhowmick@vecc.gov.in}
\email[E-mail 2: ]{29firdous11@gmail.com}
\email[E-mail 3: ]{datta@vecc.gov.in}
\email[E-mail 4: ]{dnb@vecc.gov.in}
\email[E-mail 5: ]{alok@vecc.gov.in}

\date{\today }

\begin{abstract}

    Photofission of actinides is studied in the region of nuclear excitation energies that covers the entire giant dipole resonance (GDR) region. The mass distributions of $^{238}$U photofission fragments have been explored theoretically for eight different endpoint bremsstrahlung energies from 11.5 MeV to 67.7 MeV which correspond to average photon energy of 9.09 MeV to 15.90 MeV. Among these energies, the 29.1 MeV corresponds to the average photon energy of 13.7$\pm$0.3 MeV which coincides with GDR peak for $^{238}$U photofission. The integrated yield of $^{238}$U photofission as well as charge distribution of photofission products are calculated and its role in producing nuclei and their neutron-richness is investigated.

\vskip 0.2cm

\end{abstract}

 
\maketitle

\noindent
\section{ Introduction }
\label{section1}

    Induced fission by neutron, proton and photon has been studied for several decades. However, there are still areas, e.g. far away from the valley of stability, nuclei having characteristically different structure and dynamics need to be studied. In this context the study of mass and charge distribution in low energy photon \cite{Na13,De13a,Ka15}, neutron \cite{Na13} and proton \cite{De13b} induced fission of actinides, where fission fragments are primarily neutron-rich, is found extremely important. 
    
    For producing neutron-rich radioactive ion beam by the photofission method, the use of energetic electron is a promising tool to produce energetic photons covering the entire peak of the giant dipole resonance (GDR). The beam of incident electrons of $\sim$10-50 MeV can be slowed down in a tungsten (W) converter or directly in the target (U) itself, generating bremsstrahlung photons \cite{Es80} which induces fission \cite{Di99,TRIUMF12}. Although the photofission cross section at GDR energy for $^{238}$U is about an order of magnitude lower than for the 40-MeV neutron-induced fission, still it is advantageous because the electron and $\gamma-$photon conversion efficiency is much more significant than that for the deuterons and neutrons. Moreover, at lower energies the photofission and neutron-induced fission cross sections become comparable \cite{Na13}, making the latter further disadvantageous. Therefore there is a renewed interest presently to go for photofission using e-LINAC as a primary accelerator to produce energetic photons in the GDR range. The Advanced Rare IsotopE Laboratory (ARIEL) at TRIUMF is under construction \cite{TRIUMF11}, where a superconducting electron LINAC will produce 25 MeV 100 kW e-beam for photofission (10$^{13}$ fission/s) of Uranium in the first phase.Yields of around 10$^{11}$ fission/s of Uranium by photofission have been experimentally achieved at JINR, Dubna \cite{Dubna} and at ALTO, IPN, Orsay \cite{Alto} by using 25 MeV, 20 $\mu$A e-beam.  As an extension of the present RIB development, a facility called ANURIB (Advanced National facility for Unstable and Rare Isotope Beams) will be coming up at this center with e-LINAC as primary accelerator for photofission \cite{Ch13}. 
    
    Neutron rich nuclei are generally produced by both slow neutron capture (s-process) and rapid neutron capture (r-process) processes. However, in a very high neutron flux environment, (such as during explosive events like Supernovae explosions where the temperature at the core reaches as high as 2-3$\times$10$^{9}$  $^o$K) r-process play a crucial role in heavy element synthesis through short-lived exotic nuclei. The predicted r-process reaction path has mostly been inaccessible experimentally. Using RIB one hopes to study this region in detail. A key question is whether the unusual properties of exotic nuclei alter the r-process reaction rates and path. It is therefore, important to estimate the production cross section of n-rich nuclei produced through $^{238}$U photofission and investigate how many nuclei in the r-process path can be reached in the laboratory. 

\begin{figure}[htbp]
\vspace{0.0cm}
\eject\centerline{\epsfig{file=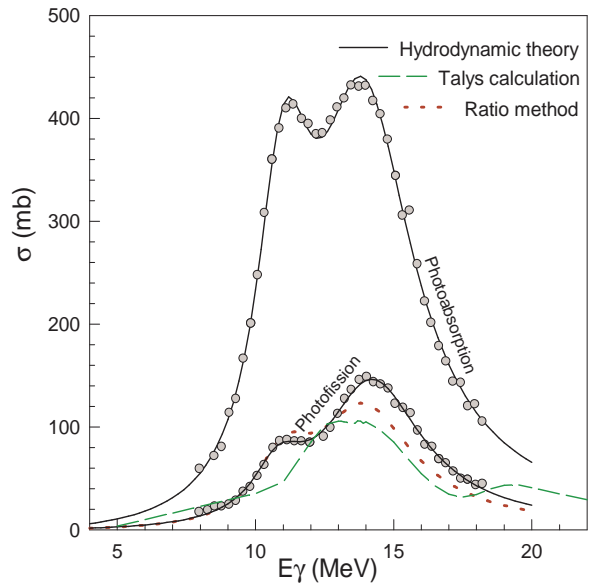,height=7.7cm,width=7.7cm}}
\caption
{Comparison of the measured photoabsorption and photofission cross sections (grey circles) \cite{Ve73} for $^{238}$U  as functions of incident photon energy with predictions (solid lines) of the hydrodynamic theory of photonuclear reactions for the giant dipole resonance region that consists of Lorentz line shapes for spherical nuclei. The dotted line represents photofission cross sections obtained by using ratio method while the dashed line represents the calculations of the TALYS code for the same.}
\label{fig1}
\vspace{0.0cm}
\end{figure}
\noindent 
    
    In the present work, we perform a simultaneous analysis and a comparison of the behavior of the symmetric and asymmetric modes \cite{Br90,Du01} of photofission induced by bremsstrahlung photons in the region of excitation energies of the $^{238}$U at the endpoint bremsstrahlung energy of 11.5$-$67.7 MeV which correspond to average photon energies of 9.09$-$15.90 MeV \cite{Na13,Is14,Be15} which includes the GDR peak for $^{238}$U photofission. A calculation of the contributions from various fission modes to the mass distribution of photofission fragments was performed in \cite{De08} at two accelerator energies. Without quantitative calculations, the existence of contributions from various fission modes was highlighted in \cite{Po93}. The results obtained in this way are compared with the predictions of the multimode-fission model for the dependence of individual fission modes on the excitation energy of the fissioning nucleus. The integrated yield of $^{238}$U photofission and charge distribution of photofission products are calculated. The role of photofission mass yield and its charge distribution in the production of neutron-rich nuclei are explored.
      
\noindent
\section{ Theoretical framework of photonuclear reaction }
\label{section2}    
  
     The total photo absorption cross section are calculated with help of hydrodynamic theory of the interaction  between photons and nuclei, where the shape of fundamental resonance in the absorption cross-section is given by  \cite{St50,Da58} 

\vspace{-0.1cm}
\begin{equation}
\sigma_a^{GDR}(E_{\gamma})= {\bf \huge\Sigma}_{i=1}^2 \frac{\sigma_i}{1+{\Big [} \frac{(E_{\gamma}^2-E_i^2)}{E_{\gamma} \Gamma_i}{\Big ]} ^2}
\label{seqn1}
\vspace{-0.1cm}
\end{equation}
\noindent
where $\sigma_i$, $E_i$ and $\Gamma_i$ are the peak cross section, resonance energy and full width at half maximum, respectively. We find that like the photoabsorption cross sections, the photofission cross sections can also be described quite well by Eq.(1). The list of parameters $\sigma_i$, $E_i$ and $\Gamma_i$ for $i=1,2$ extracted by fitting experimental data \cite{Ve73,Ca80} are provided in Ref.\cite{Bh15} for both photoabsorption as well as photofission cross sections. It was found that for photoabsorption and photofission cross sections there are little changes in parameters $E_i$ and $\Gamma_i$ while $\sigma_i$ decides the difference. Therefore, we find by fitting data of all the eight actinide nuclei that ratios $R_1=\frac{(\sigma_1)_f}{(\sigma_1)_a}$ and $R_2=\frac{(\sigma_2)_f}{(\sigma_2)_a}$ for gamma absorption and subsequent fission scale as       

\vspace{0.0cm}
\begin{equation}
 R_{i=1,2} = a_{2i}f^2+a_{1i}f+a_{0i}
\label{seqn2}
\vspace{0.0cm}
\end{equation}
\noindent
where the fissility $f=Z^2/A$ with $Z, A$ being the charge, mass numbers of the target nucleus. The scaling parameters obtained are $a_{01}=118.5486$, $a_{11}=-6.9389$, $a_{21}=0.1015$, $a_{02}=175.5418$, $a_{12}=-10.2748$ and $a_{22}=0.1504$. In Fig.1, the measured photoabsorption and photofission cross sections (grey circles) for $^{238}$U as functions of incident photon energy are compared with the predictions (solid lines) of the hydrodynamic theory of photonuclear reactions for the giant dipole resonance region that consists of Lorentz line shapes for spherical nuclei. The dotted line represents photofission cross sections obtained by using ratio method described above while the dashed line represents the calculations of the TALYS code \cite{TALYS13} for the same. It is apparent from Fig.1 that the Lorentz line shape fitting is quite accurate which reflects the fact that the uncertainties in the parameters \cite{Bh15} are quite small. The ratio method predictions for photofission are almost as good upto the first GDR peak whereas it is off by about ten percent near the second GDR peak, but still provides much better predictions than that calculated by the TALYS code which underestimates as well as fails to reproduce the shape, or by the evaporation-fission process of the compound nucleus which although retains more or less the shape of the curve but overestimates \cite{Mu10} the photofission cross sections at these energies.   
         
\noindent
\section{ Calculation of mass and charge yields in photofission  }
\label{section3}

    In the multimode-fission model, the mass distribution is interpreted as a sum of the contributions
from the symmetric and asymmetric fission modes. Each fission mode corresponds to the passage through the fission barrier of specific shape. For each fission mode, the yield is described in the form of a Gaussian function. The shape of the mass distributions can be fitted approximately by using five Gaussian functions (three fission modes) which represent the characteristics of the shell structure of the fissioning nuclei. For $^{238}$U fission, the symmetric fission mode (SM) is associated with the half of the mass of the fissioning nucleus ($A=117$ in the present case) and for the asymmetric fission modes ($ASMI, ASMII$) in addition to broad maxima at $A=138$ and $A=96$, the mass distribution exhibits narrower maxima in mass-number regions around $A=133$ and $A=101$. Thus the total yield of fragments whose mass number is $A$ is given by the expression
  
\begin{eqnarray}
 Y(A) =&& Y_{SM}(A) + Y_{ASMI}(A) + Y_{ASMII}(A)  \nonumber\\
 =&& C_{SM}\exp \Big[-\frac{(A-A_{SM})^2}{2\sigma^2_{SM}} \Big] \\
 && + C_{ASMI}\exp \Big[-\frac{(A-A_{SM}-D_{ASMI})^2}{2\sigma^2_{ASMI}} \Big] \nonumber\\
 && + C_{ASMI}\exp \Big[-\frac{(A-A_{SM}+D_{ASMI})^2}{2\sigma^2_{ASMI}} \Big] \nonumber\\
 && + C_{ASMII}\exp \Big[-\frac{(A-A_{SM}-D_{ASMII})^2}{2\sigma^2_{ASMII}} \Big] \nonumber\\ 
 && + C_{ASMII}\exp \Big[-\frac{(A-A_{SM}+D_{ASMII})^2}{2\sigma^2_{ASMII}} \Big]. \nonumber
\label{seqn3}
\end{eqnarray}
\noindent 
where the Gaussian function parameters $C_{SM}, C_{ASMI}, C_{ASMII}$ and $\sigma_{SM}, \sigma_{ASMI}, \sigma_{ASMII}$ are the amplitudes and widths, respectively, of the one symmetric ($SM$) and two asymmetric ($ASMI, ASMII$) fission modes and $A_{SM}$ is the most probable mass value for the symmetric fission mode with $A_{SM}-D_{ASMI}$ and $A_{SM}+D_{ASMI}$ being the most probable masses of a light and the complementary
heavy fragment in the $ASMI$ asymmetric fission mode while $A_{SM}-D_{ASMII}$ and $A_{SM}+D_{ASMII}$ being the most probable masses of a light and the complementary heavy fragment in the $ASMII$ asymmetric fission mode.

    The recoiling nucleus after absorbing $\gamma$ can be viewed as a compound nucleus having the same composition as the target nucleus but with the excitation energy \cite{Ba09,Mu07} 

\begin{eqnarray}
 E^*=&&m_0^{\prime}c^2-m_0 c^2 = m_0 c^2 [(1 + 2E_{\gamma}/m_0 c^2 )^{1/2} - 1] \nonumber \\
      =&&[m_0c^2(2E_{\gamma}+m_0c^2)]^{1/2}-m_0c^2
\label{seqn4}
\end{eqnarray}   
\noindent
where  $m_0$ and $m_0^{\prime}$ are the rest masses of the target nucleus before and after the photon absorption, respectively. In the present case where $m_0 c^2$ is very large compared to incident photon energy $E_{\gamma}$, the excitation energy is almost equal to the incident photon energy. In Figs.2-4, approximation by the above five Gaussian functions for the mass distribution $Y(A)$ of fragments per 100 fission events originating from $^{238}$U photofission induced by bremsstrahlung photons which correspond to average photon energies of 9.09 MeV, 13.7 MeV and 15.9 MeV are plotted and compared with experimental data \cite{Na13,Is14,Be15}.  In Fig.5, plots of the ratios of cross sections for average photon energies 9.09 MeV and 15.9 MeV to that for 13.7 MeV are shown as a function of exoticity parameter $\xi$ defined as $\frac{A-A_s}{A_d-A_s}$ where $A$, $A_s$ and $A_d$ are, respectively, the fragment mass number, mass number of stability line nucleus and mass number of neutron drip line nucleus for the same charge number $Z$ of the fragment so that $\xi=0$ for stable nucleus and $\xi=1$ for nucleus on neutron drip line. The mass number $A_d$ is obtained using mass table \cite{Au03,Au12}. The values of $A_{SM}$, $D_{ASMI}$ and $D_{ASMII}$ are 117, 21 and 16, respectively, whereas the other six parameter values for 13.7$\pm$0.3 MeV (that coincides with GDR peak for $^{238}$U photofission) are $C_{SM}=0.49\pm$0.25, $\sigma_{SM}=4.47\pm$3.33, $C_{ASMI}=5.90\pm 0.28$, $\sigma_{ASMI}=5.96\pm 0.28$, $C_{ASMII}=2.29\pm0.27$, $\sigma_{ASMII}=1.62\pm0.39$. Obviously, the fact that the uncertainties in the parameters for the symmetric mode (single Gaussian) is large while those for asymmetric modes (double Gaussians) are quite small, suggests that the importance of symmetric mode is less in determining the mass distribution. The Fig.6 shows the variations of six parameter values with the average photon energies ranging from 9.09 MeV to 15.90 MeV. The lines represent least square fits assuming quadratic energy dependence of the parameters.       

    The isobaric charge distribution of photofission products can be well simulated by a Gaussian function as
 
\begin{equation}
 Y(A,Z) = \frac{Y(A)}{\sqrt{\pi C_p}} \exp \Big[-\frac{(Z-Z_s+\Delta)^2}{C_p} \Big]
\label{seqn5}
\end{equation}
\noindent    
where  $Z_s$ represents most stable isotope of fission fragment with mass number $A$ while $\Delta$ measures the departure of the most probable isobar from the stable one. The atomic number $Z_s$ for the most stable nucleus is calculated by is differentiating the liquid drop model mass formula while keeping mass number $A$ constant and setting the term $\partial M_{nucleus}(A,Z)/\partial Z\mid_A$ equal to zero \cite{Ch06}. The values of the parameter $C_p$ which decides the dispersion and the shift parameter $\Delta$ for the most probable isotope are extracted by fitting experimental data \cite{Is14} to be 0.8 and 3.8, respectively \cite{Bh15}. In the present work the form of the charge distribution is assumed to be independent of average photon energy for the range considered here.  
    
\noindent
\section{ Results }
\label{section4}

    The photofission cross sections $\sigma_f^{GDR}$ are calculated using Lorentz line shape of Eq.(1) for gamma absorption while replacing $\sigma_i$ by the ratio method described in section-II. The production cross sections of individual fragments for $^{238}$U photofission induced by bremsstrahlung photons are obtained by multiplying fission cross section by charge distribution which means $\sigma_f(A,Z)= \sigma_f^{GDR}.Y(A,Z)/100$. The endpoint energies of 11.5$-$67.7 MeV (the energy of electrons which produce bremsstrahlung gammas when stopped by a $W$ converter) is so chosen because it corresponds to the average gamma energies of 9.09$-$15.9 MeV that covers the entire GDR range for $^{238}$U photofission. 
      
\begin{figure}[htbp]
\vspace{0.0cm}
\eject\centerline{\epsfig{file=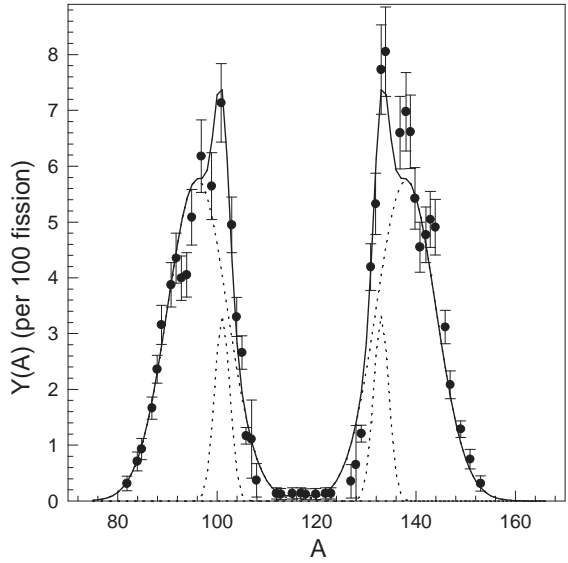,height=7.4cm,width=7.4cm}}
\caption
{Comparison of the measured mass yield distribution (full circles) \cite{Na13} for $^{238}$U photofission induced by bremsstrahlung photons having average photon energy of 9.09 MeV with the prediction (solid line) of the five Gaussian formula for yield $Y(A)$ per 100 fission events. The dotted lines represent individual contributions of one symmetric Gaussian and two asymmetric double Gaussians.}
\label{fig2}
\vspace{2.75cm}
\end{figure}
\noindent
  
\begin{figure}[htbp]
\vspace{0.0cm}
\eject\centerline{\epsfig{file=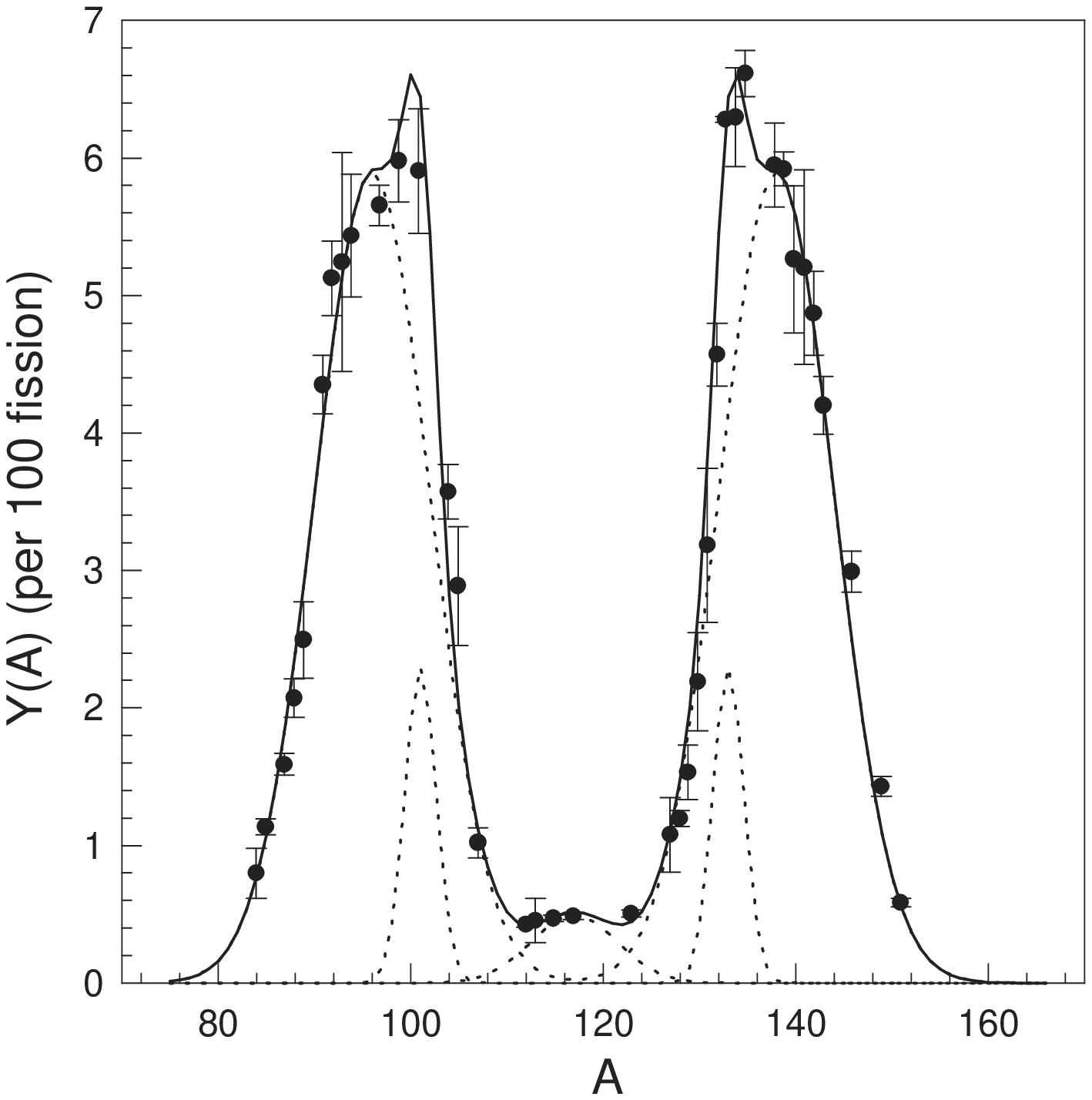,height=7.4cm,width=7.4cm}}
\caption
{Same as Fig.2 but for average photon energy of 13.7 MeV.}
\label{fig3}
\vspace{0.0cm}
\end{figure}
\noindent

\begin{figure}[htbp]
\vspace{0.0cm}
\eject\centerline{\epsfig{file=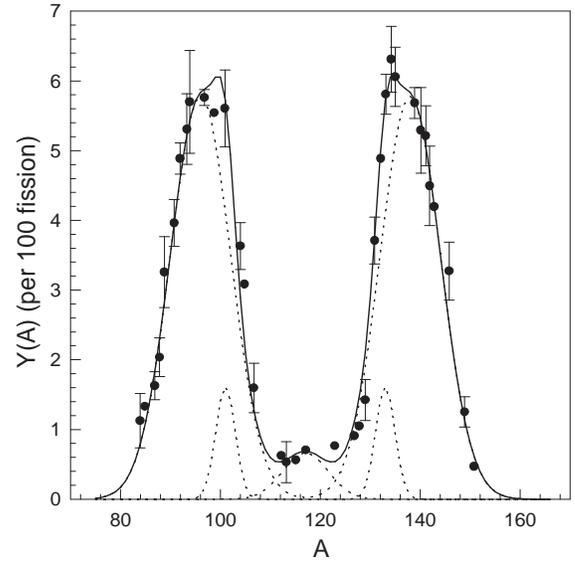,height=7.4cm,width=7.4cm}}
\caption
{Same as Fig.2 but for average photon energy of 15.9 MeV.}
\label{fig4}
\vspace{2.75cm}
\end{figure}
\noindent 
    
    It is important to mention here that with increasing energy, amplitude increases, although not so appreciable in the energy domain considered here. Furthermore, among the two asymmetric modes the peak of the $ASMII$ Gaussian is higher than the $ASMI$ and also with increasing energy the peak of the $ASMI$ Gaussian decreases while the $ASMII$ remains almost constant. The atomic numbers $Z_s$ used in Eq.(5) for the most stable nuclei are calculated using values $a_c=0.71$ MeV and $a_{sym}=23.21$ MeV \cite{Ch06}. 

    In Fig.7, production of neutron rich nuclei by the photofission of $^{238}$U by bremsstrahlung gammas with average photon energies of 15.9 MeV, 13.7 MeV and 9.09 MeV are compared. It is evident from Fig.7 that for average photon energy of 13.7$\pm$0.3 MeV (which coincides with GDR peak for $^{238}$U photofission) produces nuclei farthest from the stability line. The neutron rich species produced with cross sections $>$100 fb at energies 9.09 MeV and 15.9 MeV are 743 and 765 respectively, while that at energy 13.7 MeV is 777, the maximum. In an effort to investigate the production cross section of neutron rich nuclei by bremsstrahlung gammas, some isotopes with maximum cross sections (appearing at the two asymmetric peaks of the mass distributions), some with little lower cross sections (appearing at the symmetric mass distributions) are arranged in Table-I with average photon energies of 15.9 MeV, 13.7 MeV and 9.09 MeV. It is evident from Table-I, that with increase of energy from 9.09 MeV to 13.7 MeV the cross sections increase by about five times, whereas it decreases by a factor of two with about 2 MeV further increase in energy. Moreover, in Table-II the production cross sections of some r-process nuclei are highlighted for average photon energy of 13.7 MeV only where cross sections are found to lie in $\mu$b range: {\it e.g.} $^{80}$Zn and $^{134}$Sn, the waiting point nuclei are produced with 2.6 $\mu$b and 0.18 $\mu$b. Comparing Table-I $\&$ II, one may notice that for $A/Z$ almost equal to $2.55 \pm 0.01$, when $Z-Z_s=4$, the cross sections are of the order of mb, while for $A/Z$ almost equal to 2.66, when $Z-Z_s=6$ the cross sections reduces to $\mu$b order for fission of $^{238}$U which has $A/Z$ nearly equal to 2.58. The reduction of cross section is consistent since cross sections fall rapidly with increasing mass number because of the neutron richness which is obvious from Eq.(5). In the r-process path neutron capture stops as neutron separation energy falls approximately below 2 MeV. In our calculation, the production cross section of $^{162}$Ce is found to be 3.2$\times$10$^{-15}$ mb, where calculated value of last neutron separation energy is 2.2 MeV according to mass table \cite{Au03,Au12}. It is worthwhile to mention here that at high energies \cite{Bh98} the projectile fragment separator type RIB facilities, being developed in different laboratories, could also provide the scope for producing many new exotic nuclei through fragmentation of high energy radioactive ion beams \cite{Au95}. 

    In Fig.5, it may be noticed that with increasing exoticity $\xi$, the energy dependence remains flat irrespective of the fact whether cross sections are five times more or less by a factor of two. However, for Zr the situation is remarkably different. The increase in $\xi$ in the range 0.6-0.8 is found to be five times to twentythree times while with further increase in energy to 15.9 MeV the cross sections become half independent of $\xi$. 
    
\begin{figure}[hbp]
\vspace{0.0cm}
\eject\centerline{\epsfig{file=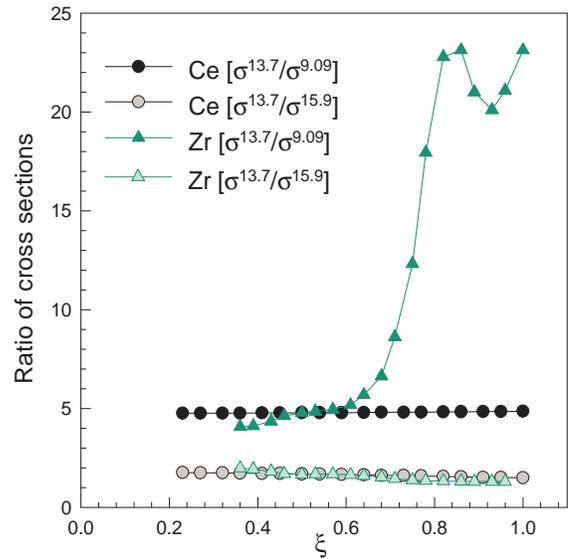,height=7.4cm,width=7.4cm}}
\caption
{Plots of ratios of cross sections for different photon energies as a function of exoticity parameter $\xi$.}
\label{fig5}
\vspace{0.0cm}
\end{figure}
\noindent
                                           
\begin{table*}[htbp]
\vspace{0.0cm}
\caption{\label{tab:table1} The the energy dependence of exotic nuclei production cross sections by photonuclear reaction in the GDR range.}
\begin{tabular}{|c|c|c|c|c|c|c|c|c|}
\hline
Elements&~~$A/Z$~~&~~$Z_s$~~&~~$A_s$~~&$~~Z_s-Z$~~&~~$A^F-A^F_s~~$ &~~~~~$\sigma^{9.09}$(mb)~~~~~&~~~~~$\sigma^{13.7}$(mb) ~~~~~&~~~~~$\sigma^{15.9}$(mb)~~~~~\\

\hline

$^{138}$Xe&   2.55&   58&  132  & 4& 6 &$8.182\times 10^{-1}$&$3.893\times 10^{0}$&$2.184\times 10^{0}$ \\ \hline

$^{133}$Te&   2.56&   56&  128  & 4& 5 &$9.147\times 10^{-1}$&$3.715\times 10^{0}$&$1.875\times 10^{0}$ \\ \hline

$^{101}$Zr&   2.53&   44&   90  & 4& 11 &$1.185\times 10^{0}$&$4.813\times 10^{0}$&$2.429\times 10^{0}$ \\ \hline

$^{96}$Sr &   2.53&   42&   88  & 4& 8 &$9.194\times 10^{-1}$&$4.375\times 10^{0}$&$2.454\times 10^{0}$ \\ \hline

$^{80}$Ge &   2.50&   36&   74  & 4& 6 &$2.407\times 10^{-2}$&$1.156\times 10^{-1}$&$7.022\times 10^{-2}$ \\ \hline

$^{117}$Pd&   2.54&   50&   106 & 4& 9 &$1.631\times 10^{-2}$&$3.280\times 10^{-1}$&$2.489\times 10^{-1}$ \\ \hline

$^{122}$Cd&   2.54&   52&   114 & 4& 8 &$2.360\times 10^{-2}$&$2.907\times 10^{-1}$&$2.083\times 10^{-1}$ \\ \hline

$^{86}$Se &   2.52&   38&    80 & 4& 6 &$2.239\times 10^{-1}$&$1.069\times 10^{0}$&$6.163\times 10^{-1}$ \\ \hline

$^{148}$Ba&   2.55&   62&   138 & 4& 10 &$2.326\times 10^{-1}$&$1.111\times 10^{0}$&$6.403\times 10^{-1}$ \\ \hline

\hline
\end{tabular} 
\vspace{2.99cm}
\end{table*}

\begin{figure}[htbp]
\vspace{0.0cm}
\eject\centerline{\epsfig{file=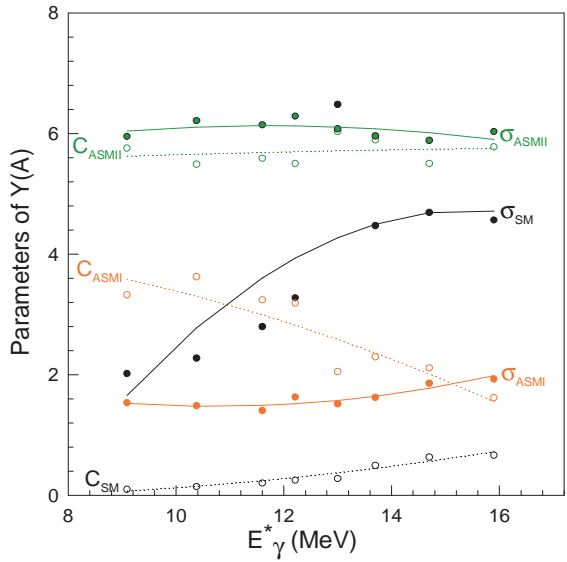,height=7.4cm,width=7.4cm}}
\caption
{Variations of the parameters of symmetric Gaussian and asymmetric double Gaussians with excitation energies.}
\label{fig6}
\vspace{0.0cm}
\end{figure}
\noindent
          
\begin{table}[htbp]
\vspace{0.0cm}
\caption{\label{tab:table1} The theoretical cross sections for production of nuclei with same A/Z for average photon energy 13.7 MeV.}
\begin{tabular}{|c|c|c|c|c|c|c|}
\hline
Elements&~$A/Z$~&~$Z_s$~&~$A_s$~&$Z_s-Z$&$A^F-A^F_s$&~~~~$\sigma$(mb)~~~~\\

\hline

$^{80}$Zn &   2.66&   36&   64  & 6&   16&  $2.6\times 10^{-3}$ \\ \hline

$^{96}$Kr &   2.66&   42&   84  & 6&   12&  $2.1\times 10^{-3}$ \\ \hline

$^{106}$Zr&   2.66&   46&   91  & 6&   15&  $3.9\times 10^{-3}$ \\ \hline

$^{133}$Sn&   2.66&   56&   119 & 6&   14&  $2.2\times 10^{-3}$ \\ \hline

$^{143}$Xe&   2.66&   60&   131 & 6&   12&  $6.6\times 10^{-3}$ \\ \hline

$^{154}$Ce&   2.66&   64&   140 & 6&   14&  $2.2\times 10^{-3}$ \\ \hline

\hline
\end{tabular} 
\vspace{0.0cm}
\end{table}        
                  
\begin{figure}[h!]
\vspace{0.0cm}
\eject\centerline{\epsfig{file=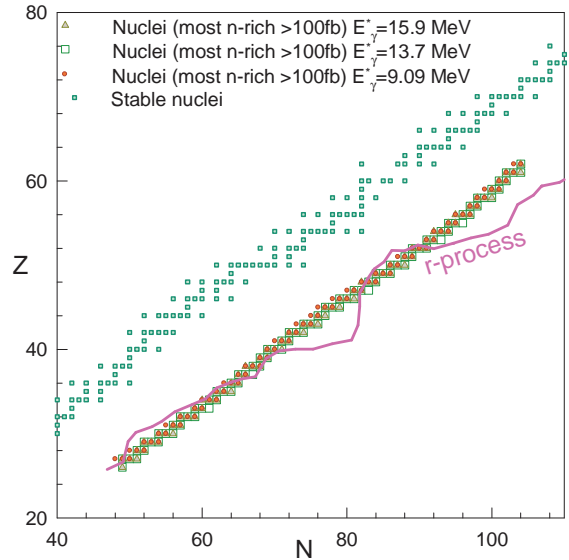,height=7.4cm,width=7.4cm}}
\caption
{Plots of atomic number $Z$ versus neutron number $N$ for exotic nuclei produced by the photofission of $^{238}$U by bremsstrahlung $\gamma$'s with average photon energy of 15.9, 13.7 and 9.09 MeV.}
\label{fig7}
\vspace{0.0cm}
\end{figure}
\noindent

\noindent
\section{ Summary and conclusion }
\label{section5}

    In summary, we find that like the photoabsorption cross sections, the photofission cross sections can also be described quite well by Lorentz line shapes. The ratio method predictions for photofission are almost as good as the Lorentz line shape fitting, whereas the evaporation-fission process of the compound nucleus overestimates the photofission cross sections. A detailed analysis of the production of each nuclear isobar via fission and the mass distributions of products originating from the photofission induced by bremsstrahlung photons is provided. The endpoint energy is chosen in the range 11.5$-$67.7 MeV, which includes of 29.1 MeV, that produce bremsstrahlung gammas when slowed down by a $W$ converter. The average gamma energy of 13.7 MeV, which coincides with GDR peak for $^{238}$U photofission, produces the maximum cross-section.

    The present calculation indicates clearly that products having $A/Z$ less than that of $A/Z$ of fissioning nucleus are more probable than otherwise. In fact, products with neutron richnesss $8 \pm 3$ have cross sections in the mb range whereas it reduces to $\mu$b order for neutron richness in the range of $14 \pm 2$. Energy dependence is significant for products having cross sections in the mb range only. However, for many of the r-process nuclei in intermediate mass range can be produced for three different energies that are very close and among them, the end point energy of 29.1 MeV is the maximum. Therefore, so far marching away from $\beta$-stability towards neutron-rich nuclei is concerned it appropriate to keep energy of the electron within 50 MeV for e-LINAC construction. For producing neutron rich nuclei, in the higher mass range, more than A$=$140, one would need to go for nuclear process other than photonuclear reaction to produce them.

\end{document}